\documentstyle[]{article}
\input epsf
\date{}
\title{The Universality of the Stellar Initial Mass Function}

\author{Paolo Padoan\thanks{Theoretical Astrophysics Center, Juliane Maries Vej 30, DK-2100 Copenhagen, Denmark}, 
\AA ke Nordlund$\thanks{Theoretical Astrophysics Center, Juliane Maries Vej 30, DK-2100 Copenhagen, Denmark 
Astronomical Observatory, Juliane Maries Vej 30, DK-2100 Copenhagen, Denmark}$, 
\& Bernard J. T. Jones\thanks{Imperial College of Science Technology and Medicine, Blackett Laboratory, Prince Consort Road, London SW7 2BZ, UK}}

\begin{document}
\maketitle
\begin{abstract}

We propose that the stellar initial mass function (IMF) is
universal in the sense that its functional form arises as a 
consequence of the statistics of random supersonic flows. 

A model is developed for the origin of the stellar IMF, that contains
a dependence on the average physical parameters (temperature, density, 
velocity dispersion) of the large scale site of star formation. The model
is based on recent numerical experiments of highly supersonic random flows
that have a strong observational counterpart.

It is shown that a Miller-Scalo like IMF is naturally produced by the model
for the typical physical conditions in molecular clouds. A more ``massive''
IMF in star bursts is also predicted.

\end{abstract}

\section{Introduction}

Star formation is a central problem of Astrophysics and Cosmology.
It is very difficult to interpret observations of galaxies, or to
predict their evolution, without any theoretical idea about the 
process of star formation.

While star formation rates and efficiencies can be constrained, thanks
to the negative feedback of the process (young massive stars are able
to disperse the star forming gas), the mass distribution of the stars is not
easily constrained theoretically. 

Moreover, observing the stellar initial mass function 
(IMF) is difficult. In old systems, most massive stars have already
evolved into cold white dwarfs and are hardly detectable. The mass of
very small and long lived stars is not easily inferred from the 
photometry. Bound stellar systems (globular clusters and open
clusters) undergo a strong dynamical evolution (mass segregation,
evaporation) that can significantly affect the IMF. In very young systems
(e.g.\ young embedded clusters), the relation between the IMF and the 
luminosity function (LF) is strongly dependent on the assumed star 
formation history (e.g.\ initial burst or continuous). Finally, the 
determination of the IMF cutoff at the smallest masses requires very 
deep stellar counts, since it may be located at masses smaller
than $0.1M_{\odot}$.

Most theoretical attempts to predict the IMF have been based on the idea
of gravitational fragmentation. This idea is a direct consequence of 
linear gravitational instability: in a system with very small density
and velocity fluctuations, gravitational instability causes the collapse
of structures larger than a critical mass, that is the mass for which
the thermal energy of the gas is comparable with its gravitational
energy. During the collapse, if cooling is efficient, the critical
mass becomes smaller, and substructures can collapse inside the collapsing
object. 

The picture of gravitational fragmentation depends on several idealized assumptions.
The collapse of substructures, and its
final result, are highly dependent on the presence of suitable 
perturbations in the density or velocity field. Moreover, fragmentation is
stopped at some point, when opacity becomes important, but this occurs
at a mass scale that depends on unknown geometrical factors, that 
affect the rate of radiative loss. Finally, the whole idea of gravitational 
fragmentation relies on linear gravitational instability, that is on the
assumption that the density field is initially almost uniform, and the
velocity field irrelevant. This is meaningful in the study of the formation
of galaxies in the Universe, since we know that the Universe is initially
very uniform. When we discuss smaller scales, instead, we are
normally dealing with a contracting background where non-uniformities 
get amplified, rather than an expanding one. On scales smaller 
than galaxies, for example on the scale of giant molecular cloud complexes
($10^5-10^6M_{\odot}$), the density and the velocity fields are highly
nonlinear and hierarchical, so that the idea of gravitational instability
cannot be applied directly, using mean values of the physical parameters.

The next level of complexity is to still maintain the idea of a critical
mass for gravitational instability, but using a distribution of the
values of the physical parameters for its definition. The distribution
of the physical parameters should be as close as possible to the actual
distribution in the star forming system, so that the complexity of
the nonlinear velocity and density field is bypassed using a statistical 
approach.

A further step is that of arguing that several ways of injecting
and transferring kinetic energy in star forming systems exist, like 
gravity, magnetic fields, fluid turbulence, supernovae and HII regions,
winds from young stars, tidal fields, thermal and magnetic instabilities,
galactic shear. All these sources of energy contribute to the generation 
of random motions, that, mediated by fluid turbulence, establish universal 
flow statistics. 

When cooling is very efficient, the most important statistic is
the density distribution. Once this is determined, the critical mass for
gravitational instability can be defined along that distribution, resulting
in a distribution of collapsing objects, or protostars. 

In the present work, which is an improvement on Padoan (1995),
we make use of recent numerical and observational
results, that allow us to describe the density distribution in 
supersonic random flows (such as the ones in molecular clouds), and
therefore to derive the mass distribution of protostars. 

Previous physical models for the origin of the stellar IMF do not include 
a description of the effect of supersonic random flows, in the star
formation sites, based on the solution of the compressible fluid equations.
Nevertheless, the idea that supersonic motions play an important part in 
the dynamics of molecular clouds, and in the formation of protostars,
has already been expressed in the literature (e.g.\ McCrea 1960; Arny 1971; 
Larson 1981; Hunter \& Fleck 1982; L\'{e}orat et al. 1990; Elmegreen 1993)

The paper is organized as follows. In the next section we present 
the numerical and observational results on the density distribution in 
molecular clouds. Section 3 contains the derivation of the protostar
mass function, whose dependence on physical parameters is discussed in
Sections 4 and 5. We then proceed to make comparisons with observations in
Section 6. The paper ends with a general discussion, followed by a summary. 
aipproc.sty 

\section{The density field in random supersonic flows}

In this section we give a statistical description of the density field that 
emerges from randomly forced supersonic flows. Such motions are 
present in dark clouds, where stars are formed. 

The statistical description is based on numerical experiments, but it is 
also confirmed by stellar extinction observations in dark clouds.

\subsection{Random supersonic flows in numerical experiments}

Nordlund and Padoan (1997) have recently discussed the importance of 
supersonic flows in shaping the density distribution in the cold interstellar medium 
(ISM).

They have run numerical simulations of isothermal flows randomly forced
to high Mach numbers. Their experiments are meant to represent a
fraction of a giant molecular cloud, where in fact such random supersonic 
motions are observed. Most details about the numerical code, which solves the
equations of magneto-hydrodynamics in three dimensions and in a supersonic regime,
and about the experiments are given in Nordlund \& Padoan; here we only 
summarize the main results.

The physical parameters of the simulated system are: $\sigma_{v}=2.5km/s$, 
$T=10K$ (therefore rms Mach number about 10), $M=4000M_{\odot}$,
$L=6pc$, where $L$ is the linear size of the periodic box.

It is found that the flow develops a complex system of interacting shocks, and
these are able to generate very large density contrasts, up to 5 orders of
magnitude, $\rho_{max}/\rho_{min}\approx 10^5$. In fact, most of the mass 
concentrates in a small fraction of the total volume of the simulation, with a very
intermittent distribution. The probability density function of the density field is
well approximated by a Log-Normal distribution:
\begin{equation}
p(lnx)dlnx=\frac{1}{(2\pi\sigma^{2})^{1/2}}exp\left[-\frac{1}{2}
\left(\frac{lnx-\overline{lnx}}{\sigma}\right)^{2} \right]dlnx
\label{1}
\end{equation}
where $x$ is the relative number density:
\begin{equation}
x=n/ \overline{n}
\label{2}
\end{equation}
and the standard deviation $\sigma$ and the mean $\overline{lnx}$ are functions
of the rms Mach number of the flow, $\cal{M}$:
\begin{equation}
\overline{lnx}=-\frac{\sigma^{2}}{2}
\label{3}
\end{equation}
and
\begin{equation}
\sigma^{2}=ln(1+{\cal{M}}^{2} \beta^2)
\label{4}
\end{equation}
or, for the linear density:
\begin{equation}
\sigma_{linear}=\beta{\cal{M}}
\label{5}
\end{equation}
where $\beta\approx0.5$.
Therefore, the standard deviation grows linearly with the rms Mach number
of the flow.

It is also found that the power spectrum, $S(k)$, of the density distribution
is consistent with a power law:
\begin{equation}
S(k)\propto k^{-2.6}
\label{6}
\end{equation}
where $k$ is the wavenumber.

The fact that the standard deviation of the linear density field, $\sigma_{linear}$,
grows linearly with the rms Mach number of the flow can be easily understood. 
The density contrast behind an isothermal shock is proportional
to ${\cal{M}}^2$, that is the square of the Mach number, but the dense shocked 
gas occupies only a fraction ${\cal{M}}^{-2}$ of the original volume. Since the
standard deviation is a volume average, the two effects result in a linear growth
of $\sigma_{linear}$ with ${\cal{M}}$. This is in fact the result, if one computes 
$\sigma_{linear}$ for the simple case of a single strong, isothermal, plane shock,
that sweeps all the mass of the system.

\subsection{Random supersonic flows in dark clouds}

An observational counterpart of the numerical experiments on supersonic
random flows has been recently indicated by Padoan, Jones, \& Nordlund
(1997), reinterpreting the observational results by Lada et al. (1994).

Lada et al. performed infrared stellar extinction measurements, 
through the dark cloud IC5146 in Cygnus. They 
obtained values of extinction for more than a thousand stars, sampled 
the observed area with a regular grid, and measured the mean and the
dispersion of the extinction determinations in each bin of the grid.
They found that the dispersion grows with the mean extinction.

This result is an indication that the absorbing material in the dark 
cloud has structure well below the resolution of the extinction map.
Padoan, Jones, \& Nordlund have shown that an intermittent 3-D
distribution in the cloud, in particular a Log-Normal distribution,
explains in a natural way the growth of dispersion with mean extinction. 
They have also shown that the observational data can be used to
constrain the value of the standard deviation and of the spectral index 
(power law power spectrum) of the 3-D density distribution. 
The observational constraints are in good agreement with the numerical 
predictions. 

Therefore, both numerical results and observations show that the random
supersonic flows, known to be present in dark clouds, result in a very intermittent
density distribution, well described by a Log-Normal statistic.
Both the values of the standard deviation and of the spectral index of such 
distribution are predicted numerically and confirmed observationally.

\section{The derivation of the stellar IMF}

A simple way to define a mass distribution of protostars is that of 
identifying each protostar with one local Jeans' mass. In this way the 
protostar MF is simply a Jeans' mass distribution. Since the gas is
cooling rapidly, the temperature is uniform, and the Jeans' mass
distribution is just determined by the density distribution.

The concept of the `local' Jeans' mass is meaningful in our scenario
for molecular clouds (MCs), because random supersonic motions (cascading from
larger scale) are present, and are responsible for shaping the density 
field. Strong density enhancements, that is to say the 
local convergence of the flow, are due to nonlinear hydro-dynamical
interactions, rather than to the local gravitational potential. We
therefore suggest a description of star formation where random motions are first 
creating a complex and highly nonlinear density field (through isothermal
shocks), and gravity then
takes over, when each `local' Jeans' mass (defined with the local density)
collapses into a protostar.

The statistic of the density field is not sufficient in general to
predict the protostar MF. Some extra knowledge on the topology of the 
density field is necessary.  

For example, the distribution of mass in a complex system 
of interacting shocks may be hierarchical. The mass distribution in 
MCs is also found to be hierarchical
over a very large range of scales (Scalo 1985; Falgarone and P\'{e}rault 1987; 
V\'{a}zquez-Semadeni 1994). This brings a considerable difficulty
when trying to define a mass distribution of cores inside MCs (Myers, Linke 
and Benson 1983; Blitz 1987; Carr 1987; Loren 1989; Stutzki and 
G\"{u}sten 1990; Lada, Bally and Stark 1991; Nozawa et al. 1991; 
Langer, Wilson and Anderson 1993; Williams and Blitz 1993), and 
indeed any mass distribution estimated from molecular emission line maps
is ill-defined, if the hierarchical structure is not taken into account. 

The main uncertainty in the Jeans' mass distribution, derived as a 
transformation of the density distribution, is related to the density
fluctuations that are smaller than their Jeans' mass. If many fluctuations
are smaller than their Jeans' mass, the transformation of density into 
Jeans' mass overestimates the number of collapsing protostars with
that mass. 

Nevertheless, in our numerical experiments we find that isolated density 
fluctuations, smaller than their Jeans mass, are extremely rare, and do not 
account for more than $1\%$ in mass, for any level of density considered.
We conclude therefore that the transformation of the density field into 
the distribution of Jeans' masses gives an estimate of the mass
distribution of collapsing objects, that is not in error by more than a 
few percent.

The density distribution per unit volume is given by equation (\ref{1}). 
If we multiply that function with the relative density $x$, we get the 
density distribution per unit mass, that is the mass fraction at any given 
density:

\begin{equation}
f(x)dx=xp(x)dx
\label{7}
\end{equation}

The fraction of the total mass in collapsing structures of mass $<M$ 
is the integral of the distribution $f(x)$ over relative densities
$x > x_{J}$:

\begin{displaymath}
\int_{x_{J}}^{\infty}f(x)dx
\end{displaymath}
where $x_{J}$ is the Jeans' density for the mass $M$. The Jeans' mass 
distribution is the derivative along mass of the previous integral:

\begin{equation}
F(M_{J})=f(x_{J})\frac{dx_{J}}{dM_{J}}
\label{10}
\end{equation}

The Jeans' mass can be written as:

\begin{equation}
M=M_{J}=1M_{\odot}Bx^{-1/2}
\label{11}
\end{equation}
where:

\begin{equation}
B=1.2\left(\frac{T}{10 K}\right)^{3/2}\left(\frac{\overline{n}}{1000 cm^{-3}}\right)^{-1/2}
\label{12}
\end{equation}
is the average Jeans' mass, that is the Jeans' mass for the average relative
density $x=1$.

Here we use the simplest definition of Jeans' mass: without turbulent
pressure or rotation, because the gas has just been shocked and is dissipating
its kinetic energy in a short time; without magnetic pressure;
we will discuss the role of the magnetic field in such random flows
in a subsequent work (Padoan \& Nordlund, 1997).

Using equations (\ref{1}), (\ref{7}), (\ref{10}), (\ref{11}), and (\ref{12}) we get the 
protostar MF:

\begin{eqnarray}
& F(M)dM= \frac{2B^2}{(2\pi\sigma^2)^{0.5}}M^{-3} \nonumber \\
 & exp\left[-\frac{1}{2}\left(\frac{2lnM-A}{\sigma}\right)^2\right]dM
\label{13}
\end{eqnarray}
where $M$ is in solar masses, and:

\begin{equation}
A=2lnB-\overline{lnx}
\label{14}
\end{equation}

One can also express the MF in average Jeans' mass, instead of
in solar masses:
\begin{eqnarray}
& F(M/B)d(M/B)= \frac{2}{(2\pi\sigma^2)^{0.5}}\left(\frac{M}{B}\right)^{-3} \nonumber \\
  &exp\left[-\frac{1}{2}\left(\frac{2ln(M/B)-|\overline{lnx}|}{\sigma}\right)^2\right]d(M/B) \nonumber
\end{eqnarray}

A linear plot of the protostar MF is shown in Fig.~1, for $T=10K$. 
One recognizes
a long tail at large masses and an exponential cutoff at the smallest masses,
inherited from the Log-Normal distribution of density.
This shape is an important result, because most models for the origin
of the stellar IMF are not able to reproduce the cutoff at the smallest masses,
which should be present in any reasonable IMF.

In the coming sections we will discuss the dependence of the MF on the average
physical parameters of the star forming gas, and we will then compare
our results with the observations.

\section{The dependence of the IMF on the physical parameters}

The protostar MF depends on the density distribution that arises
from random supersonic motions, through a complex system of interacting 
shocks, and on the definition of the Jeans' mass. The first dependence 
brings into the MF the dependence on the average temperature, $T$, and 
velocity dispersion of the flow, $\sigma_{v}$, through the rms Mach number
of the flow, which is the only parameter of the density distribution in random 
supersonic flows. The dependence on the Jeans' mass translates into a 
dependence of the MF on the average density, $n$, and on the temperature.
Therefore our model for the MF may be applied to different sites of
star formation, identified by their mean values of density, temperature
and velocity dispersion.

In Figs.~2, 3, and 4, we have plotted mass distributions for different values of the 
physical parameters. We have chosen to plot the exponent of the power
law approximation of the MF, rather than the actual MF.  The exponent is
defined as:
\begin{equation}
X=\frac{\partial ln(F(lnM))}{\partial lnM}=\left(\frac{2A}{\sigma^2}-3\right)-
\frac{4}{\sigma^2}lnM
\label{15}
\end{equation}
The Salpeter MF has $X=-1.35$, and the Miller-Scalo MF (Miller \& Scalo 1979) has $X=-1.0-0.43lnM$,
where $M$ is in solar masses. 

The most probable stellar mass per logarithmic mass interval, that is
the stellar mass that contributes most to the MF, is defined by
$X(M_{max})\equiv 0$, along the curves $X(M)$ plotted in the figures.

In Fig.~2 we see that a growing $T$ produces a flattening of
the MF at large masses, and a growth of $M_{max}$, 
that is the typical stellar mass. The effect of the growth of
density is illustrated in Fig.~3; its effect
is the opposite of the effect of the temperature. Fig.~4 shows the dependence 
on velocity dispersion. 

Note that, although the effect of temperature and density is 
qualitatively as 
expected from the definition of the Jeans' mass, the effect of $T$
on the MF is more complicated than through the Jeans' mass, because $T$ 
also affects the density distribution, through the Mach number. 

We will now discuss the variation of the position of
the cutoff (or of the typical stellar mass) with the physical parameters.

\section{The typical stellar mass}

The main result of a theory of star formation should be the prediction
of the typical stellar mass. From this point of view, all models
resulting in a power law MF are unsuccessful, because power
laws are featureless.

In the present work we have shown that the random supersonic motions 
present in molecular clouds produce a protostar MF with an
exponential cutoff at the smallest masses, just below the
most probable protostellar mass, $M_{max}$.

The position of the maximum in the MF is given by imposing
\begin{displaymath}
X(M_{max})\equiv 0
\end{displaymath}
in equation (\ref{15}). By using the definition of $A$ from (\ref{14}),
of $\sigma$ from (\ref{4}), and of $M_{J}$ from (\ref{11}) and (\ref{12}), 
we get:
\begin{equation}
M_{max}= 1M_{\odot}Be^{\left(-\frac{1}{2}\sigma^2\right)}
\label{16}
\end{equation}
where $1M_{\odot}B$ is the average Jeans' mass, that is the Jeans' mass
for the average density. Therefore:
\begin{equation}
M_{max}= 0.2M_{\odot}\left(\frac{n}{1000cm^{-3}}\right)^{-1/2}
\left(\frac{T}{10 K}\right)^{2}\left(\frac{\sigma_{v}}{2.5 km/s}\right)^{-1}
\label{17}
\end{equation}

We can read the result as a modified Jeans' mass. The modification
is quite important. In fact this modified Jeans' mass is more sensitive
to temperature than the traditional Jeans' mass, and is also quite 
sensitive to the velocity dispersion $\sigma_{v}$. 

This result is not surprising, because it looks like the Jeans' mass
at constant external pressure (Spitzer, 1978, p.~241), 
if turbulent ram pressure is considered. 
Nevertheless it is an important result because it has been obtained
from a realistic statistical description of random supersonic flows, that allows
the prediction of the whole shape of the MF. 

Another way to interpret the modified Jeans' mass is to use equation 
(\ref{16}), and substitute the standard deviation of the logarithmic
density distribution, $\sigma$, with the linear standard deviation,
$\sigma_{linear}$, from equations (\ref{4}) and (\ref{5}). We obtain:
\begin{equation}
M_{max}= \frac{1M_{\odot}B}{\sigma_{linear}}
\label{18}
\end{equation}
where $\sigma_{linear}$ is about one half of the rms Mach number of the 
flow (cf equation (\ref{5})), and $B$ is the Jeans' mass for the mean 
density, in solar masses. Therefore we may conclude that {\it the most
probable Jeans' mass is equal to the Jeans' mass for the mean density 
divided by half of the rms Mach number}. As an example, a typical molecular 
cloud with rms Mach number about 10, and Jeans' mass of the mean density
about $1M_{\odot}$, has a most probable Jeans' mass of $0.2 M_{\odot}$.

In Fig.~5 , we show contours of constant $M_{max}$, on the plane
$n-T$, for $\sigma_{v}=3 km/s$, typical of molecular clouds.

\section{The observed IMF}

In Fig.~6 the theoretical MF for $T=10K$ (dashed line) is compared with the 
Miller-Scalo MF (MSMF) (dotted line). The shape of the theoretical MF is 
different from the shape of the MSMF. In fact, for the typical parameters of
MCs, or of MC cores, the MF is always less broad than the MSMF.
On the other hand, the models with low $T$ (say $5K$) give the correct slope 
for low masses, while the models with high $T$ (say $40K$) give the correct
slope for the large masses. 

Therefore the MSMF can be reproduced only if the solar neighborhood
stars are assumed to be born in clouds with temperatures in the range
$5-40K$, which is a reasonable assumption, since these temperature values
are measured in cloud cores. It is likely that the solar neighborhood
stars are a mixed population coming from different cloud cores, or even from 
different giant molecular cloud complexes, with temperatures in the range
observed in present day molecular clouds.

To illustrate the origin of a MF that contains a mixed population, coming
from clouds with different temperatures, we integrate our 
theoretical MF along a temperature distribution, $g(T)dT$:
\begin{displaymath}
F_{mixed}(M)dM=\int_{T} F(M,T)dMg(T)dT
\end{displaymath}   

Fig.~6 shows the result of the temperature integration (continuous line). The 
temperature distribution has been taken to be $g(T)\propto T^{-1}$, which
means that there are more cold clouds than warm ones. 

One can see that the temperature integration improves the shape of the 
single temperature MF, making the theoretical MF practically coincident
with the MSMF.

We may therefore conclude that the model is consistent with the MSMF, as long
as most of the solar neighborhood stars are formed in molecular clouds
similar to the ones that are the sites of present day star formation, 
with temperatures between $5K$ and $40K$. 

It has been claimed by many authors, on both theoretical and observational
grounds, that the IMF in star-burst regions is more ``massive'' than in the solar
neighborhood. Models of the stellar populations
in star-bursts suggest a MF with the low-mass cutoff at a few solar masses
(e.g.\ Augarde \& Lequeux 1985; Doane \& Mthews 1993; Riecke et al. 1993; 
Doyon, Joseph, \& Wright 1994).

These ``massive'' MFs are in agreement with our theoretical prediction. In fact,
a value of $4M_{\odot}$ is predicted for the cutoff in the MF, for $T\approx60K$, 
which is reasonable in environments with strong UV and X-ray radiation fields, 
and with enhanced (even by a factor 100) cosmic ray flux. We also predict
a slope of the MF considerably smaller than in the MSMF. For example,
$X=-0.9$ (the value found by Malumuth and Heap (1994) in the core of 30 Doradus, 
R136a, which is a local example of a star-burst event ) is predicted by the model 
for $T=60K$, or slightly warmer (fig.2).

Padoan, Jimenez, \& Jones (1997) have studied the hypothesis of a primordial
origin of GCs, by applying the present model of star formation to
protoglobular clouds of a few $10^8M_{\odot}$ in baryons. In their model the GCs 
originate from the star formation process in the core of the large cloud, at density
$n\approx 10^4cm^{-3}$ and temperature $T\approx 100K$ (due to $H_2$ cooling),
while most of the halo stars are the stars formed in the rest of the 
protoglobular cloud, which does not result in a bound system. For the 
halo stars the parameters are $n=250cm^{-3}$ and $T=100K$. The assumed
velocity dispersion is some fraction of the virial velocity, 
$\sigma_{v}\approx 50km/s$.

It is found that the GC MF matches the Miller-Scalo MF very well. In particular, 
the exponent of the MF, in the interval
$[0.1, 0.6]M_{\odot}$, is $X=[0.5,-0.5]$, in agreement with the most recent
results on the MF in NGC 6397 (Paresce, De Marchi, \& Romaniello 1995; 
d'Antona \& Mazzitelli 1996), and in contrast with previous results (Fahlman et al. 1989;
Richer et al. 1990, 1991).

\section{Discussion}

A purely statistical description of the origin of the IMF has been given by several 
authors, as an attempt to model the process of gravitational fragmentation
(e.g.\ Auluck \& Kothari 1954; Kruszewski 1961; Kiang 1966; Reddish 1962, 1966;
Fowler \& Hoyle 1963; Belserene 1970; Larson 1972; Elmegreen \& Mathieu 1983;
Zinnecker 1984, Di Fazio 1986).

Other works tried to relate the MF of molecular cloud cores with the stellar MF
(see Zinnecker (1993) for a discussion). 

Most of the physical models for the origin of the IMF have been based on the 
concept of opacity-limited fragmentation (Silk 1977a, 1977b, Yoshii
\& Saio 1985, 1986). 

In all cited models, the presence of random supersonic motions in the clouds is
not taken into account. Models where such motions are considered in a 
semi-empirical way are Myers \& Fuller (1993), and Silk (1995). The effect of 
turbulent motions are also considered by Arny (1971), as a source of internal 
pressure, and by Hunter \& Fleck (1982), as a source of compression.

Takebe, Unno, \& Hatanaka (1962) assumed, as in the present work, that the 
mass function is set by the distribution of the Jeans' mass in the cloud, but,
missing a physical model for the cloud structure, they inferred this from 
the IMF.

In the present work, as in Padoan (1995), we suggest that all stars
are formed as a consequence of turbulent fragmentation, that is the
fragmentation due to a complex system of strong interacting shocks, formed
in a field of random supersonic motions. 

Such an approach has been made possible only very recently, thanks to new 
numerical simulations of 3-D highly supersonic magneto-hydrodynamic random flows,
(Nordlund \& Padoan, 1997) such as the ones observed in molecular clouds, 
and to the recognition of their observational counterpart (Padoan, Jones, \& 
Nordlund 1997; Padoan \& Nordlund, 1997).

We have seen that the theoretical MF depends on the values of temperature,
density and velocity dispersion, averaged over the large scale star forming
system. Since the velocity dispersion and the density can be very different,
when measured at different scales, the application of the present model might 
seem ambiguous: how do we define the scale over which to perform the average 
of the physical parameters? 

This, in fact, is not a problem, because one is likely to find the same IMF when the 
average is performed on different scales. The reason is that in general the 
star forming gas has a hierarchical structure, such that the average density on 
large scale is smaller than the density on small scales. The velocity, instead,
grows with the scale.

As an example we consider the well known scaling relations for the ISM
(Larson, 1979; Larson, 1981; Leung, Kutner, \& Mead 1982; Myers, 1983; Quiroga, 1983; 
Sanders, Scoville, \& Solomon, 1985; 
Dame et al., 1986; Falgarone \& P\'{e}rault 1987; Fuller and Myers 1992), which
are approximately:

\begin{displaymath}
n(L)\propto L^{-1}
\end{displaymath}
where $N(L)$ is the density averaged on the linear scale $L$, and:
\begin{displaymath}
\sigma_{v}(L)\propto L^{1/2}
\end{displaymath}
where $\sigma_{v}(L)$ is the velocity dispersion averaged on the scale $L$.

We can now see how the typical stellar mass changes, when the average is
performed on different scales. Equation (\ref{17}) becomes:
\begin{equation}
M_{max}\approx 0.1 M_{\odot}\left(\frac{T}{10K}\right)^2
\label{19}
\end{equation}
that is the dependence on velocity dispersion and density cancel each other.

Therefore, the scalings found in the ISM are such that, performing the average on a
molecular cloud core of $10M_{\odot}$, or on the whole giant molecular cloud
complex of $10^6M_{\odot}$, gives a prediction, for the typical stellar mass, that is
the same. Nevertheless, the shape 
of the distribution is affected by the choice of the velocity dispersion, in the sense that
the MF becomes broader when the velocity dispersion is increased. Therefore some
care must be taken when estimating the velocity dispersion of the star formation site.

Note that the scaling (\ref{19}) does not apply to primordial clouds where globular
clusters are formed, however. The turbulent ram pressure is there much 
larger than in molecular clouds, but since the temperature is also larger
(H$_{2}$ cools the gas down to ~100 K), a characteristic stellar mass close to the 
one in molecular clouds is obtained.

As we discussed above, the protostar MF can be obtained directly from the 
density distribution of the gas, because most high density structures, formed
in the supersonic random flow, are larger than their Jeans' mass. This does not
imply that the expected star formation efficiency of molecular clouds should be 
close to $100\%$. In fact, it takes about two dynamical times before the Log-Normal
density distribution is achieved, that is about $10^7$ years, for a typical molecular
cloud of $10^5 {\rm M}_{\odot}$. It is therefore possible that the first supernova
explosions are able to disrupt the clouds before a large fraction of protostars
are formed.

Moreover, it is well known that not all the gas that starts to collapse into a protostar
will finally accrete onto the star; some fraction of it will be expelled by stellar winds.
Although we have not gone into these details in the formulation of the MF, it is clear
that this process alone can further reduce the star formation efficiency.

Finally, some parts of the cloud are strongly
magnetized, and therefore their collapse may be hindered, or delayed, until the cloud is
disrupted by supernova explosions. This influence of magnetic fields is not discussed 
here, but it is the subject of another work (Padoan \& Nordlund, 1997).

\section{Summary and Conclusions}

In the present work we have proposed a new physical model for the origin
of the stellar IMF. The model is based on a new statistical description
of star formation on large scale, that focuses on the importance of random 
supersonic flows, observed in the sites of star formation.

Recent numerical and observational results, concerning the density
distribution that arises from random supersonic motions, are implemented
in the theoretical model for the MF of protostars. The main results of the 
present work are the following:

\begin{itemize}

\item The MF is quantified without free parameters, with its dependence 
on the mean temperature, density and velocity dispersion of the star 
forming gas.

\item The shape of the protostar MF has a single maximum, a long tail of 
massive stars, and an exponential cutoff below the maximum. Such a shape 
is inherited directly from the density distribution in random supersonic flows.

\item The typical protostellar mass is 
$M_{max}\approx 0.2M_{\odot}\left(\frac{n}{1000cm^{-3}}\right)^{-1/2}
\left(\frac{T}{10 K}\right)^{2}\left(\frac{\sigma_{v}}{2.5 km/s}\right)^{-1}$, and
$M_{max}\approx 0.1 M_{\odot}\left(\frac{T}{10K}\right)^2$, using the ISM 
scaling laws.

\item A Miller-Scalo IMF is predicted for the solar neighborhood stars, if 
they are formed in molecular clouds, similar to the ones observed in 
the sites of present day star formation, with temperatures in the range 
$5K-40K$.

\item Globular clusters are expected to have a MF similar to the 
Miller-Scalo, with a typical stellar mass of $0.1M_{\odot}$.

\item Star-burst regions should have flatter IMFs, with a more massive cutoff, 
because of their higher mean temperature.

\end{itemize}

\section*{Acknowledgements}

This work has been supported by the Danish National Research Foundation
through its establishment of the Theoretical Astrophysics Center.
P. P. thanks Raul Jimenez, John Peacock, Derek Ward-Thompson and 
Francesco Lucchin for stimulating discussions.

Computing resources were provided by the Danish National Science Research Council,
and by the French `Centre National de Calcul Parall\`{e}le en Science de la Terre'.

We are grateful to the referee, for having stimulated an improved 
version of the paper.

\newpage
\begin{figure}
\centering
\leavevmode
\epsfxsize=1.0
\columnwidth
\epsfbox{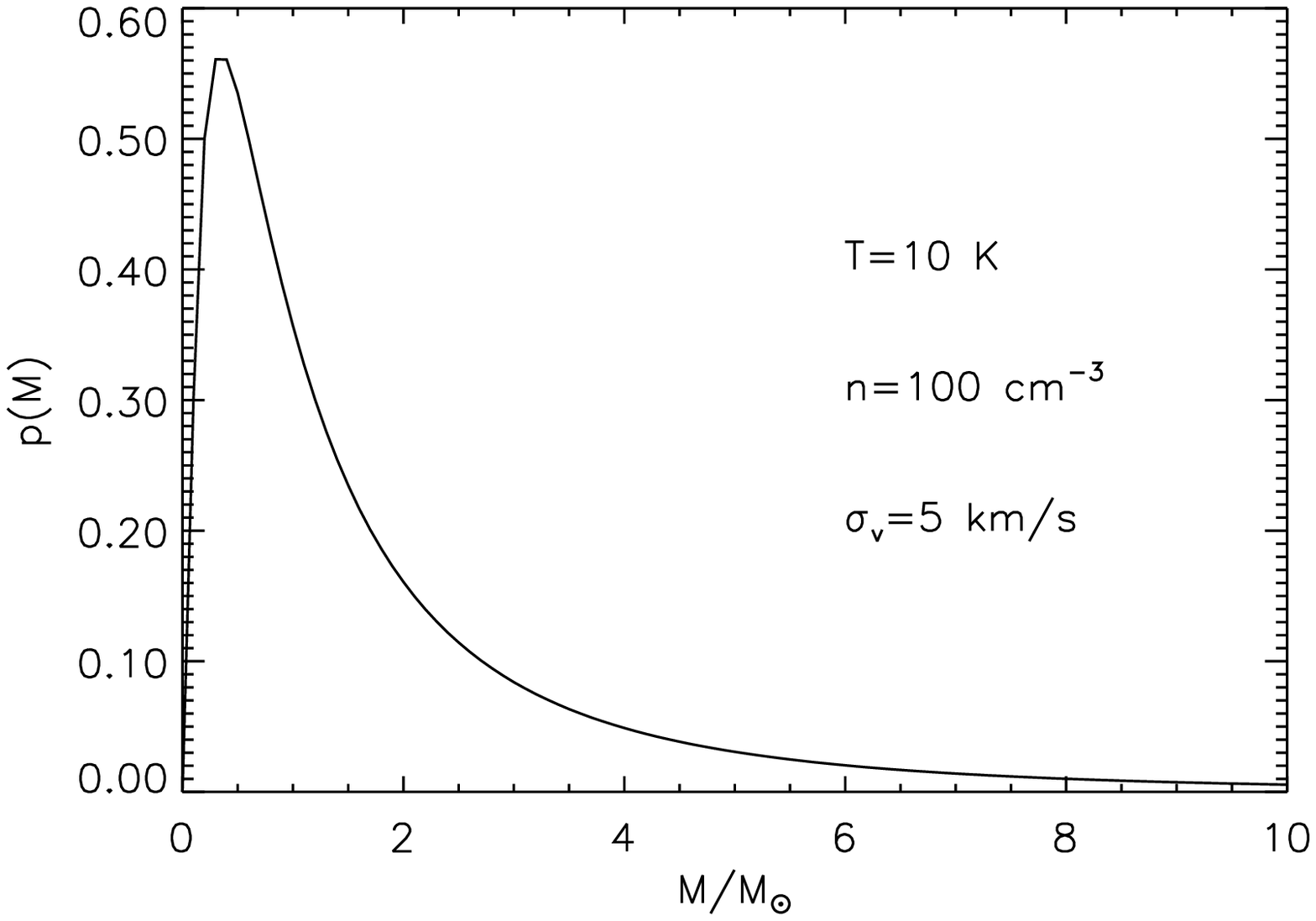}
\caption[]{A linear plot of the theoretical MF. The linear shape is characterized
by a maximum, with an exponential cutoff for smaller masses. The exponential
cutoff is an important feature, because could be identified in the observations,
without ambiguities due to uncertainties in the mass-luminosity relation.}
\end{figure}

\newpage
\begin{figure}
\centering
\leavevmode
\epsfxsize=1.0
\columnwidth
\epsfbox{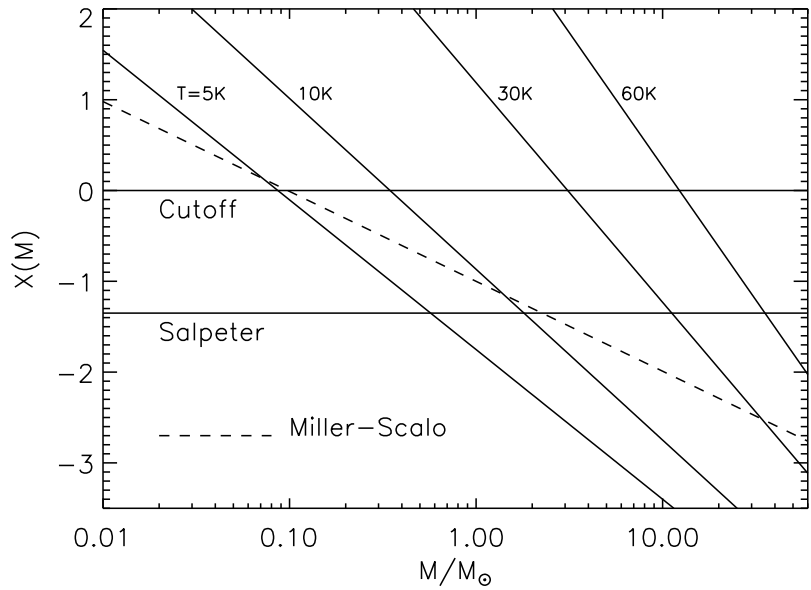}
\caption[]{The power law exponent of the theoretical MF is plotted versus the mass,
for different temperatures.
The Miller-Scalo MF (dashed line) is also plotted for comparison. The Salpeter's
value $X=-1.35$ and the cutoff value $X=0$ are also shown. The Miller-Scalo
exponent is fitted by low temperatures at low masses, and by high temperature at 
large masses. The mean density and velocity dispersion have been taken to be
$n=1000cm^{-3}$, and $\sigma_{v}=2.5km/s$, typical of molecular cloud cores.}
\end{figure}

\newpage
\begin{figure}
\centering
\leavevmode
\epsfxsize=1.0
\columnwidth
\epsfbox{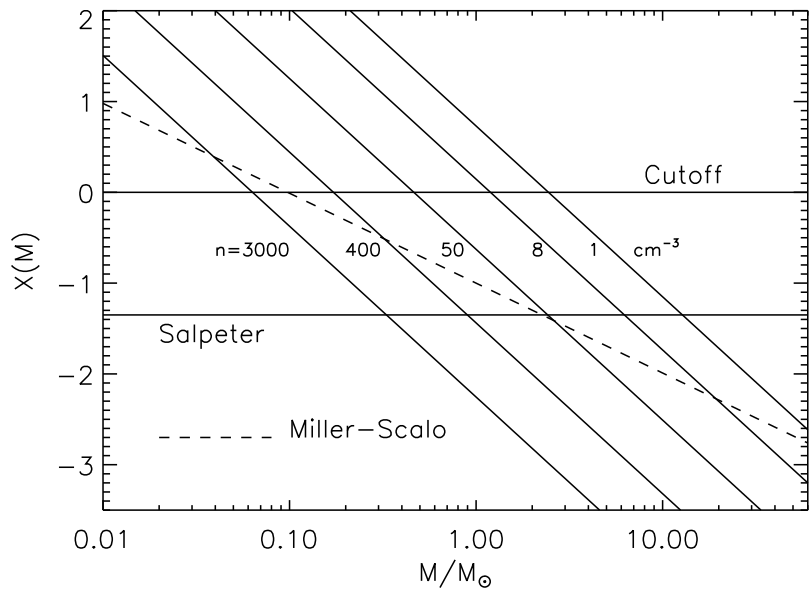}
\caption[]{The same as in Fig.~2, but for different values of density. The temperature 
and velocity dispersion have been taken to be $T=10K$, and $\sigma_{v}=2.5km/s$.
The exponent $X(M)$ varies with mass always faster than in the Miller-Scalo, 
which is an indication the the Miller-Scalo emerges from a mixed population of stars,
formed in clouds with different temperatures.}
\end{figure}

\newpage
\begin{figure}
\centering
\leavevmode
\epsfxsize=1.0
\columnwidth
\epsfbox{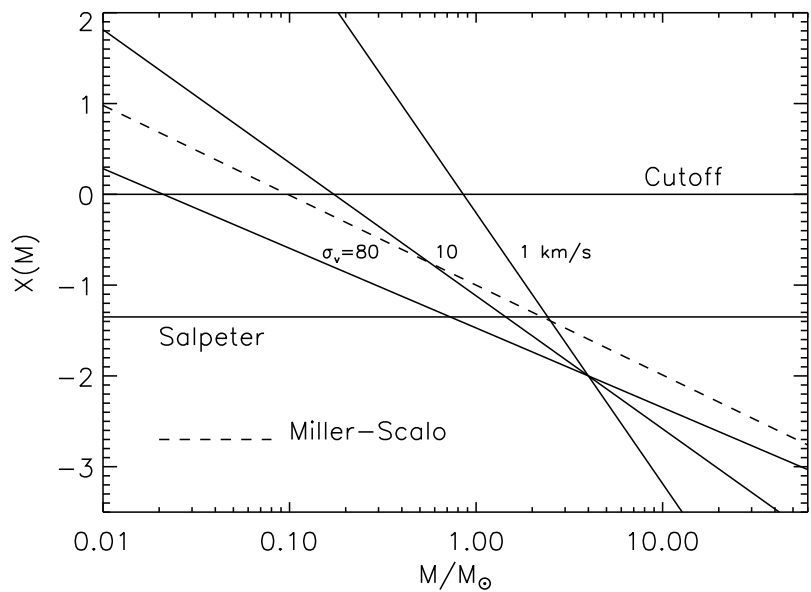}
\caption[]{The same as in Fig.~2, but for different velocity dispersions. The temperature
and density have been taken to be $T=10K$, and $n=1000cm^{-3}$. Very large velocity 
dispersions, probably typical of large primordial clouds (protogalaxies and protoglobular
clouds) can fit very well the Miller-Scalo, even for a single temperature.}
\end{figure}

\newpage
\begin{figure}
\centering
\leavevmode
\epsfxsize=1.0
\columnwidth
\epsfbox{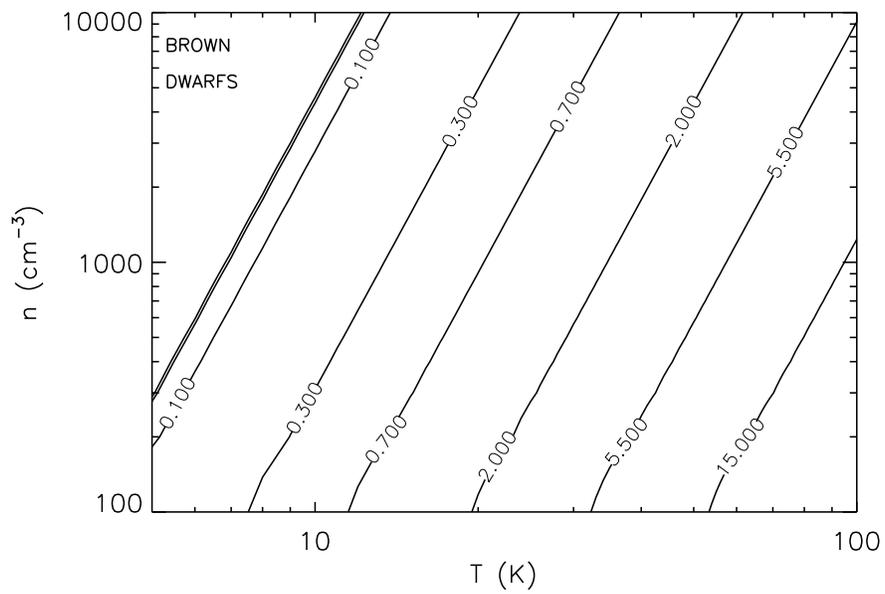}
\caption[]{Lines of constant value of the cutoff, or typical stellar mass, in solar masses,
on the plane density-temperature. The velocity dispersion is $\sigma_{v}=3.0km/s$,
typical of molecular clouds.}
\end{figure}

\newpage
\begin{figure}
\centering
\leavevmode
\epsfxsize=1.0
\columnwidth
\epsfbox{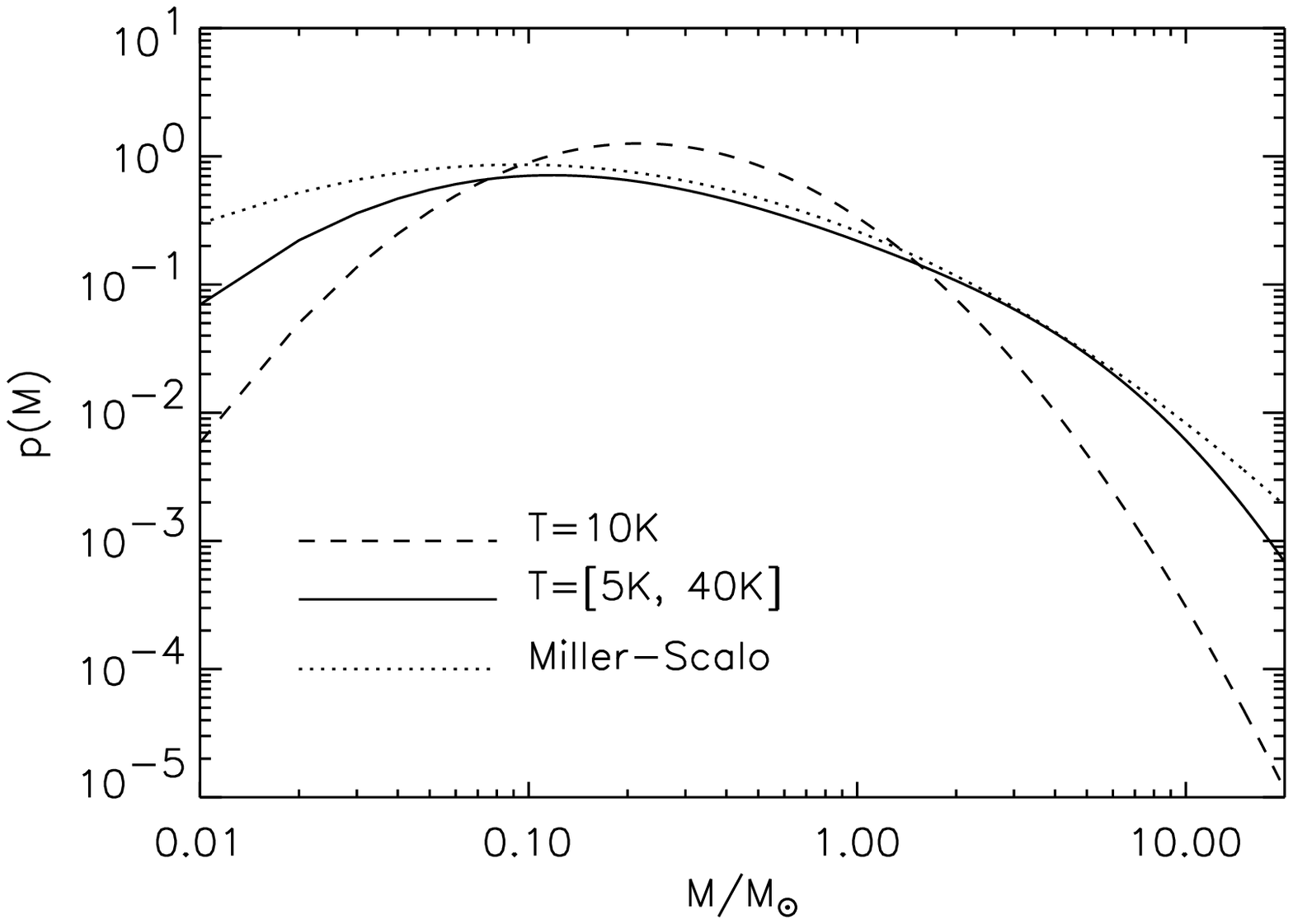}
\caption[]{Log-log plot of the theoretical MF (dashed line), for a temperature $T=10K$.
The Miller-Scalo MF is plotted for comparison (dotted line). The single temperature
MF cannot be made to fit the Miller-Scalo. Once the theoretical MF is generated
from a distribution of temperatures, in the range $5-40K$ (continuous line), it is
practically coincident with the Miller-Scalo.}
\end{figure}


\begin{thebibliography}{}

\bibitem[]{} Arny, T. 1971, ApJ, 169, 289

\bibitem[]{} Augarde, R., \& Lequeux, J. 1985, A\&A, 147, 273

\bibitem[]{} Auluck, F. C., \& Kothari, D. S. 1954, Nature, 174, 565

\bibitem[]{} Belserene, E. P. 1970, Observatory, 90, 239

\bibitem[]{} Blitz, L. 1987, in Physical Processes in Interstellar Clouds, ed. Morfill, G. E.
                   \& Scholer, M. (Reidel Publ. Comp.), 35

\bibitem[]{} Carr, J. S. 1987, ApJ, 323, 170

\bibitem[]{} Dame, T. M., Elmegreen, B. G., Cohen, R. S., \& Thaddeus, P. 1986, 
                   ApJ, 305, 892

\bibitem[]{} d'Antona, F., Mazzitelli, I. 1996, ApJ, 456, 329

\bibitem[]{} Di Fazio, A. 1986, A. \& A., 159, 49

\bibitem[]{} Doane, J. S., \& Mathews, W. G. 1993, ApJ, 419, 573

\bibitem[]{} Doyon, R., Joseph, R. D., \& Wright, G. S. 1994, ApJ, 421, 101

\bibitem[]{} Elmegreen, B. G., \& Mathieu, R. D. 1983, MNRAS, 203, 305

\bibitem[]{} Elmegreen, B. G. 1993, ApJ, 419, L29

\bibitem[]{} Fahlman, G. G., Richer, H. B., Searle, L., \& Thompson, E. B. 1989,
                   ApJ, 343, L49

\bibitem[]{} Falgarone, E., \& P\'{e}rault, M. 1987, in Physical Processes in Interstellar Clouds,
                   ed. Morfil, G. E. \& Scholer, M. (Dordrecht: Reidel), 59

\bibitem[]{} Fowler, W. A., \& Hoyle, F. 1963, Roy. Obs. Bull., 11, 1

\bibitem[]{} Fuller, G. A., \& Myers, P. C. 1992, ApJ, 384, 523

\bibitem[]{} Hunter, J. H. Jr, \& Fleck, R. C. Jr 1982, ApJ, 256, 505

\bibitem[]{} Kiang, T. 1966, Zs. Ap., 64, 426

\bibitem[]{} Kruszewski, A. 1961, Acta Astr., 11, 199

\bibitem[]{} Lada, E. A., Bally, J., \& Stark, A. A. 1991, ApJ, 368, 432

\bibitem[]{} Lada, C. J., Lada, E. A., Clemens, D. P., \& Bally, J. 1994, ApJ, 429, 694

\bibitem[]{} Langer, W. D., Wilson, R. W., \& Anderson, C. H. 1993, ApJ, 408, L45

\bibitem[]{} Larson, R. B. 1972, Nature, 236, 21

\bibitem[]{} Larson, R. B. 1979, MNRAS, 186, 479

\bibitem[]{} Larson, R. B. 1981, MNRAS, 194, 809

\bibitem[]{} L\'{e}orat, J., Passot, T., Pouquet, A. 1990, MNRAS, 243, 293

\bibitem[]{} Leung, C. M., Kutner, M. L., \& Mead, K. N. 1982, ApJ, 262, 583

\bibitem[]{} Loren, R. B. 1989, ApJ, 338, 902

\bibitem[]{} Malumuth, E. M., \& Heap, S. R. 1994, AJ, 107, 1054

\bibitem[]{} McCrea, W. H. 1960, Proc. R. Soc. Lond. A, 256, 245 

\bibitem[]{} Miller, G. E, \& Scalo, J. M. 1979, ApJS, 41, 413

\bibitem[]{} Myers, P. C. 1983, ApJ, 270, 105

\bibitem[]{} Myers, P. C., Linke, R. A., \& Benson, P. J 1983, ApJ, 264, 517

\bibitem[]{} Myers, P. C., \& Fuller, G. A. 1993, ApJ, 402, 635

\bibitem[]{} Nordlund, \AA., \& Padoan, P. 1997, in preparation

\bibitem[]{} Nozawa, S., Mizuno, A., Teshima, Y., Ogawa, A., \& Fukui, Y. 1991, ApJS,
                   77, 647 

\bibitem[]{} Padoan, P. 1995, MNRAS, 277, 377

\bibitem[]{} Padoan, P., Jones, B. J. T., \& Nordlund, \AA. 1997, ApJ, 474, 730

\bibitem[]{} Padoan, P., Jimenez, R., \& Jones, B. J. T. 1997, MNRAS, 285, 711

\bibitem[]{} Padoan, P. \& Nordlund, \AA. 1997, in preparation

\bibitem[]{} Paresce, F., De Marchi, G., \& Romaniello, M. 1995, ApJ, 440, 216

\bibitem[]{} Quiroga, R. J. 1983, Ap. Sp. Sci., 93, 37

\bibitem[]{} Reddish, V. C. 1962, Sci. Progr., 50, 235

\bibitem[]{} Reddish, V. C. 1966, in Vistas in Astronomy, Vol. 7, ed. A. Beer
                   (Oxford: Pergamon), 173

\bibitem[]{} Richer, H. B., Fahlman, G. G., Buonanno, R., \& Fusi Pecci, F.  1990, 
                   ApJ, 359, L11

\bibitem[]{} Richer, H. B., Fahlman, G. G., Buonanno, R., Fusi Pecci, F., Searle, L., \&
                   Thompson, I. B. 1991, ApJ, 381, 147

\bibitem[]{} Rieke, G. H., Loken, K., Rieke, M. J., \& Tamblyn, P. 1993, ApJ, 412, 99  

\bibitem[]{} Sanders, D. B., Scoville, N. Z., \& Solomon, P. M. 1985, ApJ, 289, 372

\bibitem[]{} Scalo, J. M. 1985, in Protostars and Planets II, ed. D. C. Black and M. S.
                  Mathews (Tucson: University of Arizona Press), 349

\bibitem[]{} Silk, J. 1977a, ApJ, 214, 152

\bibitem[]{} Silk, J. 1977b, ApJ, 214, 718 

\bibitem[]{} Silk, J. 1995, ApJ, 438, L41

\bibitem[]{} Spitzer, L., Jr. 1978, Physical processes in the interstellar medium (Wiley)

\bibitem[]{} Stutzki, J., \& G\"{u}sten,R. 1990, ApJ, 356, 513

\bibitem[]{} Takebe, H., Unno, W., \& Hatanaka, T. 1962, Pub. Astr. Soc. Japan, 14, 340

\bibitem[]{} V\'{a}zquez-Semadeni, E. 1994, ApJ, 423, 681

\bibitem[]{} Williams, J. P., \& Blitz, L. 1993, ApJ, 405, L75

\bibitem[]{} Yoshii, Y., \& Saio, H. 1985, ApJ, 295, 521

\bibitem[]{} Yoshii, Y., \& Saio, H. 1986, ApJ, 301, 587 

\bibitem[]{} Zinnecker, H. 1984, MNRAS, 210, 43 

\bibitem[]{} Zinnecker, H. 1993, in Protostars and Planets III, ed.s E. H. Levy \& J. I. Lunine
 (Tucson: University of Arizona Press), 429


\end{thebibliography}
\end{document}